\documentclass[12pt,epsf]{article}
\usepackage{graphicx,amsmath,amssymb}
\begin{document}

\def\a{\alpha}
\def\b{\beta}
\def\c{\varepsilon}
\def\d{\delta}
\def\e{\epsilon}
\def\f{\phi}
\def\g{\gamma}
\def\h{\theta}
\def\k{\kappa}
\def\l{\lambda}
\def\m{\mu}
\def\n{\nu}
\def\p{\psi}
\def\q{\partial}
\def\r{\rho}
\def\s{\sigma}
\def\t{\tau}
\def\u{\upsilon}
\def\v{\varphi}
\def\w{\omega}
\def\x{\xi}
\def\y{\eta}
\def\z{\zeta}
\def\D{\Delta}
\def\G{\Gamma}
\def\H{\Theta}
\def\L{\Lambda}
\def\F{\phi}
\def\P{\Psi}
\def\S{\Sigma}

\def\o{\over}
\def\beq{\begin{eqnarray}}
\def\eeq{\end{eqnarray}}
\newcommand{\gsim}{ \mathop{}_{\textstyle \sim}^{\textstyle >} }
\newcommand{\lsim}{ \mathop{}_{\textstyle \sim}^{\textstyle <} }
\newcommand{\vev}[1]{ \left\langle {#1} \right\rangle }
\newcommand{\bra}[1]{ \langle {#1} | }
\newcommand{\ket}[1]{ | {#1} \rangle }
\newcommand{\EV}{ {\rm eV} }
\newcommand{\KEV}{ {\rm keV} }
\newcommand{\MEV}{ {\rm MeV} }
\newcommand{\GEV}{ {\rm GeV} }
\newcommand{\TEV}{ {\rm TeV} }
\def\diag{\mathop{\rm diag}\nolimits}
\def\Spin{\mathop{\rm Spin}}
\def\SO{\mathop{\rm SO}}
\def\O{\mathop{\rm O}}
\def\SU{\mathop{\rm SU}}
\def\U{\mathop{\rm U}}
\def\Sp{\mathop{\rm Sp}}
\def\SL{\mathop{\rm SL}}
\def\tr{\mathop{\rm tr}}

\def\IJMP{Int.~J.~Mod.~Phys. }
\def\MPL{Mod.~Phys.~Lett. }
\def\NP{Nucl.~Phys. }
\def\PL{Phys.~Lett. }
\def\PR{Phys.~Rev. }
\def\PRL{Phys.~Rev.~Lett. }
\def\PTP{Prog.~Theor.~Phys. }
\def\ZP{Z.~Phys. }

\newcommand{\bea}{\begin{eqnarray}}   
\newcommand{\eea}{\end{eqnarray}}
\newcommand{\bear}{\begin{array}}  
\newcommand {\eear}{\end{array}}
\newcommand{\bef}{\begin{figure}}  
\newcommand {\eef}{\end{figure}}
\newcommand{\bec}{\begin{center}}  
\newcommand {\eec}{\end{center}}
\newcommand{\non}{\nonumber}  
\newcommand {\eqn}[1]{\beq {#1}\eeq}
\newcommand{\la}{\left\langle}  
\newcommand{\ra}{\right\rangle}
\newcommand{\ds}{\displaystyle}
\def\SEC#1{Sec.~\ref{#1}}
\def\FIG#1{Fig.~\ref{#1}}
\def\EQ#1{Eq.~(\ref{#1})}
\def\EQS#1{Eqs.~(\ref{#1})}
\def\REF#1{(\ref{#1})}
\def\TEV#1{10^{#1}{\rm\,TeV}}
\def\GEV#1{10^{#1}{\rm\,GeV}}
\def\MEV#1{10^{#1}{\rm\,MeV}}
\def\KEV#1{10^{#1}{\rm\,keV}}
\def\lrf#1#2{ \left(\frac{#1}{#2}\right)}
\def\lrfp#1#2#3{ \left(\frac{#1}{#2} \right)^{#3}}
\newcommand{\rem}[1]{{\bf #1}}


\baselineskip 0.7cm

\begin{titlepage}

\begin{flushright}
UT-11-45 \\
December, 2011
\end{flushright}

\vskip 1.35cm
\begin{center}
{\large \bf 
  Wino LSP detection in the light of recent Higgs searches\\
 at the LHC
}
\vskip 1.2cm
Takeo Moroi
and
Kazunori Nakayama

\vskip 0.4cm

{\it Department of Physics, University of Tokyo, Tokyo 113-0033, Japan}\\

\vskip 1.5cm

\abstract{ 

  Recent LHC data showed excesses of Higgs-like signals at the Higgs
  mass of around 125\,GeV.  This may indicate supersymmetric models
  with relatively heavy scalar fermions to enhance the Higgs mass.
  The desired mass spectrum is realized in the anomaly-mediated
  supersymmetry breaking model, in which the Wino can naturally be the
  lightest superparticle (LSP).  We discuss possibilities for
  confirming such a scenario, particularly detecting signals from Wino
  LSP at direct detection experiments, indirect searches at neutrino
  telescopes and at the LHC.

}

\end{center}
\end{titlepage}

\setcounter{page}{2}


Higgs mass contains very important information about low-energy
supersymmetry (SUSY) models, which is well motivated because it
provides a viable candidate of dark matter (DM) and also because it
realizes the gauge coupling unification.  In particular, in the
minimal SUSY standard model (MSSM), the lightest Higgs boson cannot be
heavier than the $Z$-boson at the tree level, while a sizable
radiative correction may enhance the Higgs mass ~\cite{Okada:1990vk}.
The size of the radiative correction depends on the masses (and other
parameters) of superparticles.  The lightest Higgs mass becomes larger
as superparticles (in particular, stops) become heavier.  Thus, once
the lightest Higgs mass is known, mass scale of superparticles is
constrained.

Recently, the ATLAS collaboration reported $3.6\sigma$ local excess of
the standard model (SM) Higgs-like event at $m_h\simeq 126\ {\rm GeV}$
\cite{ATLAS-CONF-2011-163}.  In addition, the CMS collaboration also
showed more than $2\sigma$ local excess at $m_h\simeq 124\ {\rm GeV}$
\cite{HIG-11-032}.\footnote{The excesses based on global
  probabilities, which take account of the look-elsewhere effect, are
  $2.3\sigma$ (ATLAS) and $1.9\sigma$ (CMS).}  In order to achieve
such a value of the lightest Higgs mass in the MSSM, relatively large
values of the superparticle masses are required; the typical scale of
the sfermion masses to realize $m_h\simeq 125$\,GeV is
$10$\,TeV--$10^3$\,TeV~\cite{hep-ph/0408240,arXiv:1108.6077}.  Then,
if the masses of all the superparticles are of the same order, it is
difficult to find experimental signals of low-energy SUSY and the
existence of SUSY is hardly confirmed.

Although the sfermion masses are much larger than the electroweak
scale, gauginos may be much lighter than sfermions and within the
reach of collier and other experiments.  One interesting possibility
is the model in which the SUSY breaking scalar masses are from direct
coupling to the SUSY breaking field while the gaugino masses are
generated by the anomaly-mediation
mechanism~\cite{hep-th/9810155,hep-ph/9810442}; in this letter, we
call such a model as anomaly-mediated SUSY breaking (AMSB) model.
Even in the AMSB model, however, if the pure anomaly-mediation
relation holds among the gaugino masses, gluino mass is about 8 times
larger than the mass of Wino.  Thus, if the Wino mass is a few hundred
GeV, which is the lower bound on it from astrophysical and
cosmological considerations as will be reviewed later, the gluino mass
becomes multi-TeV; with such a heavy gluino, the discovery of the SUSY
signal at the LHC becomes challenging because we consider the case
that all the squarks are extremely heavy.

Even so, there still exist possibilities of discovering signals of the
AMSB scenario.  In particular, in the present framework, the neutral
Wino is the lightest superparticle (LSP) and may be DM.  In such a
case, pair annihilation cross section of the LSP and the scattering
cross section of the LSP off the nuclei are both enhanced compared to
the Bino LSP case, which has significant implications to direct and
indirect detection of DM.  Because the search of the superparticles at
the LHC may be difficult, it is important to pursue these
possibilities and explore how well we can study the AMSB scenario with
these procedures.\footnote{
The heavy SUSY particle spectrum and their detectability were 
discussed in a different context in Ref.~\cite{Hall:2011jd}.
}

In this letter, motivated by the recent Higgs searches at the LHC, we discuss the detectability of the
signals of AMSB scenario.  We pay particular attention to the case of
the Wino LSP.  We focus on direct/indirect detection of the Wino DM at
underground laboratories and neutrino telescopes.  We also comment on
the LHC reach for the direct Wino production.  Since superparticles
except for gauginos are heavy, standard methods for SUSY searches may
not work.  Even in this case, we will show that there are some windows
for the confirmation of the SUSY.

Let us first briefly discuss important properties of the AMSB
scenario.  We assume that the soft SUSY breaking scalar masses are
generated by the direct coupling between the scalars and the SUSY
breaking hidden sector field, while the gaugino masses are generated
by the anomaly mediation mechanism.  Adopting the pure AMSB relation,
the gaugino masses are given by \cite{hep-th/9810155, hep-ph/9810442}
\begin{eqnarray}
  M_a^{\rm (AMSB)} = \frac{b_a}{16\pi^2} g_a^2 m_{3/2},
\end{eqnarray}
where $g_a$ ($a=1$--$3$) are gauge coupling constants of the SM gauge
groups, $m_{3/2}$ is the gravitino mass, and
$(b_1,b_2,b_3)=(11,1,-3)$.  Then, the Wino becomes the lightest among
the gauginos, and gaugino masses largely separate: $m_{\tilde{B}}:
m_{\tilde{W}}: m_{\tilde{g}}\simeq 3:1:8$.  Although the AMSB relation
may be affected by Higgs and Higgsino loop diagrams
\cite{hep-ph/9810442, Gherghetta:1999sw}, we adopt the pure AMSB mass
relation.  With the gaugino masses being of $O(100)\ {\rm
  GeV}$--$O(1)\ {\rm TeV}$, the gravitino mass becomes of $O(10)\
{\rm TeV}$--$O(100)\ {\rm TeV}$.  The sfermion masses are expected to
be of the same order of the gravitino mass, which is preferred from
the point of view of realizing $m_h\simeq 125\ {\rm GeV}$.  In
particular, if the scalar masses are (almost) equal to the gravitino
mass, $m_h\simeq 125\ {\rm GeV}$ requires relatively small value of
$\tan\beta\sim$ a few (where $\tan\beta$ is the ratio of the vacuum
expectation values of up- and down-type Higgs bosons)
\cite{arXiv:1108.6077}.

Before discussing the detectability of the signals of AMSB model, we
comment on the supersymmetric Higgs mass parameter (so-called
$\mu$-parameter).  In the present setup, the soft SUSY breaking scalar
mass parameters of up- and down-type Higgs bosons are expected to be
of $O(10)\ {\rm TeV}$--$O(100)\ {\rm TeV}$.  In order to have viable
electroweak symmetry breaking, the $\mu$-parameter (as well as heavy
Higgs boson masses) is also expected to be of the same order; then,
the Higgsinos become extremely heavy and the Wino becomes the LSP.
Thus, we pay particular attention to the case of Wino LSP in the
following.  In some of our following analysis, however, we consider
the case with $\mu\sim O(100)\ {\rm GeV}$--$O(1)\ {\rm TeV}$ taking
account of the possibility of an accidental tuning of the parameters.
This is because detection rates of some of signals (in particular, the
direct detection rates) strongly depend on the value of $\mu$.

Taking account of the radiative correction due to the gauge boson
loops, the neutral Wino becomes lighter than the charged one.
Therefore, we focus on the case of neutral Wino LSP.  In addition, we
assume that the LSP (i.e., the neutral Wino) is the dominant component
of DM.  The Wino LSP accounts for the present DM density for
$m_{\tilde W} \simeq 3$\,TeV if it is produced only from thermal
bath~\cite{hep-ph/0610249}.  In the AMSB scenario, however, the Wino
LSP can be non-thermally produced from the gravitino or moduli decay
\cite{hep-ph/9810442, hep-ph/9906527}.  If the reheating temperature
takes an appropriate value, for example, the decay of gravitino
produces the Wino LSP with correct relic density
\cite{hep-ph/0012052}, while thermal leptogenesis~\cite{Fukugita:1986hr} works
successfully~\cite{Ibe:2011aa}.  Thus the Wino is a good DM
candidate in the present setup.  Hereafter, we assume that the right
amount of Wino is somehow produced in the early universe to be DM.

\begin{figure}[tbp]
\begin{center}
\includegraphics[scale=0.5]{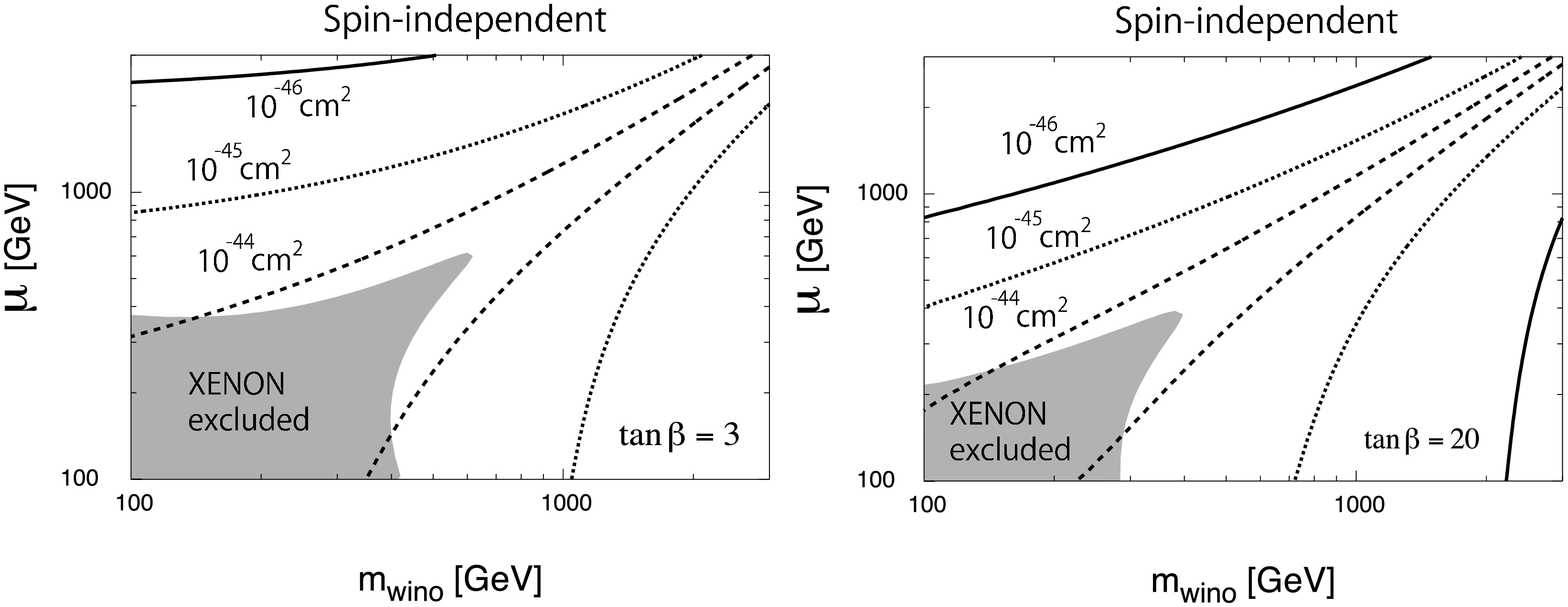}
\vskip 1cm
\includegraphics[scale=0.5]{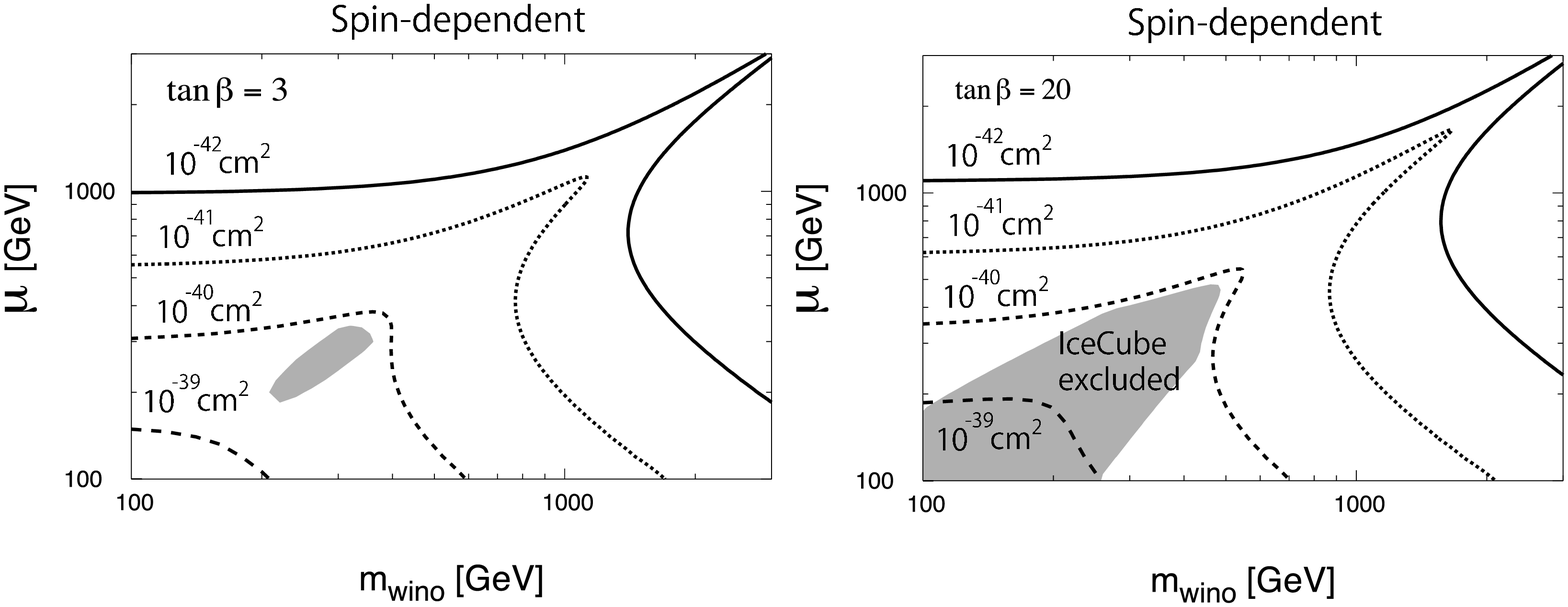}
\caption{
	Contours of spin-independent (SI) and spin-dependent (SD) Wino-proton scattering cross sections
	are plotted on the plane of $m_{\rm wino}$ and $\mu$.
	Shaded regions are excluded by the XENON100 experiment for SI,
	and IceCube experiment for SD.
 }
\label{fig:SI}
\end{center}
\end{figure}

We start with discussing direct detection experiments of DM.  The
scattering cross section of the Wino LSP off the nucleon significantly
depends on $\mu$.  Since all scalars except for the lightest Higgs
boson are expected to be heavy enough, it is only the lightest Higgs
boson that mediates the spin-independent (SI) scattering.  The
DM-proton scattering cross section is given by~\cite{Jungman:1995df}
\begin{equation}
    \sigma = \frac{4}{\pi} 
    \left( \frac{m_{\tilde \chi^0} m_N}{m_{\tilde \chi^0} + m_N} \right)^2 
    \left[ \left( n_p f_p + n_n f_n \right)^2 
    + 4 \frac{J+1}{J} 
    \left( a_p \langle s_p \rangle + a_n \langle s_n \rangle \right)^2    
  \right], 
\end{equation}
where the first and the second term in the bracket are the
contributions of SI and spin-dependent (SD) interaction, respectively.
Here $m_{\tilde \chi_0}$ is the LSP mass, $m_N$ is the mass of the
target nucleus, $n_p (n_n)$ is the number of proton (neutron) in the
target nucleus, $J$ is the total nuclear spin, $a_p$ and $a_n$ are the
effective DM-nucleon SD couplings, and $\langle s_{p(n)} \rangle$ are
the expectation values of the spin content of the proton and neutron
groups within the nucleus.  The effective DM-proton coupling, $f_p$,
is given by
\begin{equation}
    f_{p} = \sum_{q=u, d, s} \frac{f_q^{H}}{m_q} m_p f_{T_q}^{(p)} 
    + \frac{2}{27}f_{T_G} \sum_{q=c, b, t} \frac{f_q^{H}}{m_q} m_p,
\end{equation}
where $f_{T_G}=1-\sum_{u,d,s}f_{T_q}^{(p)}$, $m_p$ and $m_q$ denote
the proton and quark masses, respectively, and $f_q^H$ is the
effective DM-quark coupling obtained by the exchange of the Higgs
boson.  Since the DM-Higgs coupling is proportional to the magnitude
of Wino-Higgsino mixing,\footnote{ In the limit of pure Wino DM, the
  Wino-nucleon scattering cross section is too small to be
  detected~\cite{arXiv:1004.4090}.  } the cross section is enhanced if
the Wino-Higgsino mixing is large.  In Fig.~\ref{fig:SI} we plot the
Wino-proton SI and SD scattering cross section.  In this plot we have
used following values for the quark contents in the
proton~\cite{Ohki:2008ff} : $f_{T_u}^{(p)}=0.023, f_{T_d}^{(p)}=0.034,
f_{T_s}^{(p)}=0.025$ and taken $\tan \beta = 3$ and $\tan\beta=20$.
The XENON100 experiment~\cite{arXiv:1104.2549} most severely
constrains the SI cross section.  The sensitivity is improved by a few
orders of magnitude for the next generation 1 ton scale detectors, and
then broad parameter regions up to $m_{\tilde W}\sim \mu \sim 1$\,TeV
will be explored.  The IceCube searches for neutrino events arising
from the DM annihilation in the Sun.  Since the efficiency for the DM
trapping into the Sun depends on the DM-proton scattering cross
section, the high-energy neutrino observations give limits on it.  For
the SD cross section, the IceCube gives the most stringent limit, and
it will be further improved by about one order of magnitude with the
DeepCore instrument~\cite{arXiv:1111.2738}.

We have also calculated the detection rate at the IceCube DeepCore,
arising from high-energy neutrinos produced by the Wino annihilation
at the Galactic Center (GC).  We distinguish two event classes
following Refs.~\cite{arXiv:0911.5188,arXiv:0912.3521,arXiv:1009.2068}
: contained muon events and shower events.  The contained muons
correspond to those emerge inside the instrumental volume through the
high-energy neutrino interactions with nucleons.  The shower events
are caused by charged current interactions of electron and tau
neutrinos, and neutral current interactions of all neutrino species.
They leave electromagnetic/hadronic shower inside the instrumental
volume.  The event rate of the contained muons is given by
\begin{equation}
	N_{\mu^+\mu^-} = \int dE_{\nu_\mu}\int_{E_{\rm th}}^{E_{\nu_\mu}} dE_\mu
	\left[ \frac{d\Phi_{\nu_\mu}}{dE_{\nu_\mu}}
		\left( \frac{d\sigma_{\nu_\mu p}^{\rm (CC)}}{dE_\mu}n_p
		+\frac{d\sigma_{\nu_\mu n}^{\rm (CC)}}{dE_\mu}n_n \right)
		+(\nu_\mu \leftrightarrow \bar \nu_\mu)
	\right] V_{\rm eff}(E_\mu),   \label{cont}
\end{equation}
where $E_{\nu_\mu}$ is the incident neutrino energy, $E_\mu$ is the
muon energy resulting from the neutrino-proton (neutron) interactions,
$E_{\rm th}$ is the threshold energy above which the muon can be
detected, $d\Phi_{\nu_\mu}/dE_{\nu_\mu}$ is the neutrino flux at the
Earth, $d\sigma_{\nu_\mu p(n)}^{\rm (CC)}/dE_\mu$ denotes the
neutrino-proton (neutron) charged current cross section for producing
the muon energy with $E_\mu$, $n_p (n_n)$ is the proton (neutron)
number density in the detector material, and $V_{\rm eff}$ is the
effective volume for the muon detection.  The incident neutrino flux
generated by DM annihilation from the GC within cone half angle of
$\theta$ is given by
\begin{equation}
  \frac{d\Phi_{\nu_i}}{dE_{\nu_i}} = 
  \frac{R_{\odot}\rho_{\odot}^2}{8\pi m_{\tilde W}^2}
  \left( \sum_{j=e,\mu,\tau} \langle \sigma v\rangle 
    \frac{dN_{\nu_j}}{dE_{\nu_j}}P_{j\to i} \right)
  \langle J_2 \rangle_\Omega \Delta \Omega.
\end{equation}
Here $R_\odot = 8.5$\,kpc and $\rho_\odot=0.3\,{\rm GeV}\,{\rm
  cm^{-3}}$, $\langle \sigma v\rangle$ is the Wino self-annihilation
cross section including the non-perturbative
effect~\cite{hep-ph/0307216}, and $dN_{\nu_j}/dE_{\nu_j}$ is the
energy spectrum of the neutrino produced by DM annihilation, which is
calculated by the PYTHIA package for the $WW$ final
state~\cite{hep-ph/0603175}, $P_{j\to i}$ is the probability that the
$\nu_j$ at the production is converted to $\nu_i$ because of the
neutrino oscillation effect, $\Delta\Omega=2\pi(1-\cos\theta)$, and
$\langle J_2 \rangle_\Omega$ includes the information about the DM
density profile in the Galaxy~\cite{arXiv:0812.0219}.  The shower
event is evaluated in a similar way to the contained muon events
(\ref{cont}), except that the charged current interactions from
$\nu_e$ and $\nu_\tau$ as well as the neutral current interactions for
all neutrino flavors are included.  The background event is evaluated
by inserting the atmospheric neutrino flux into the expression
(\ref{cont}).

\begin{figure}[tbp]
\begin{center}
\includegraphics[scale=1.7]{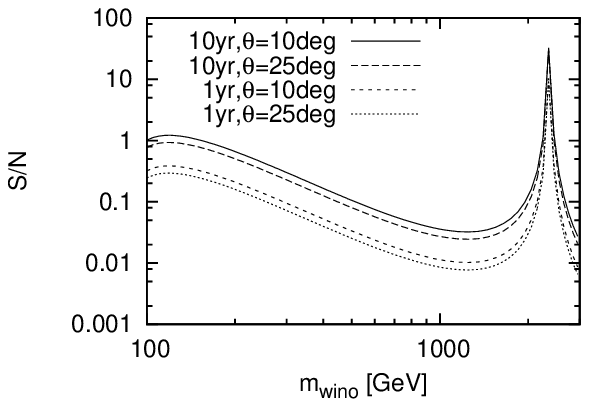}
\vskip 1cm
\includegraphics[scale=1.7]{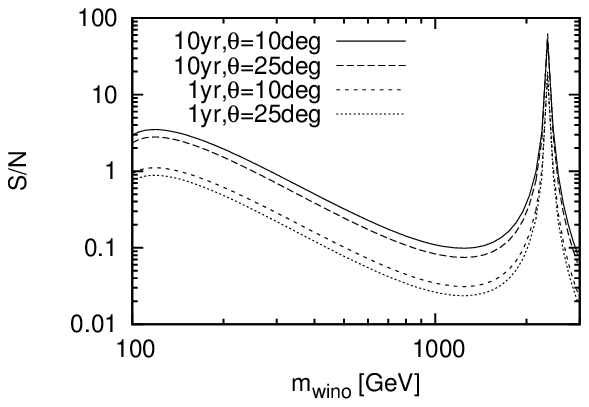}
\caption{
  Signal-to-noise ratio at the IceCube DeepCore as a function of the Wino mass.
  Sensitivities for contained muon events (upper panel) and shower events
  (lower panel)
  with 1 year and 10 year observations are shown.
  We have considered the neutrino flux from the cone half angle 
  $\theta = 10^{\circ}$ and $\theta = 25^{\circ}$ around the GC.
}
\label{fig:ICDC}
\end{center}
\end{figure}

Fig.~\ref{fig:ICDC} shows the signal-to-noise ratio at the IceCube
DeepCore as a function of the Wino mass.  Sensitivities for contained
muon events (upper panel) and shower events (lower panel) with 1 year
and 10 year observations are shown.  We have adopted the NFW density
profile and considered the neutrino flux from the cone half angle
$\theta = 10^{\circ}$ and $\theta = 25^{\circ}$ around the GC.  As
noted in Ref.~\cite{arXiv:1009.2068}, the sensitivity is maximized for
$\theta \simeq 10^{\circ}$.  For this cone half angle, the flux
dependence on the DM density profile is not
large~\cite{arXiv:0812.0219}.  The effective volume for the contained
and shower events are set to be $0.04\,{\rm km^3}$ and $0.02\,{\rm
  km^3}$, respectively~\cite{arXiv:0911.5188}.  The atmospheric
background is taken from Ref.~\cite{astro-ph/0611418}.  It is seen
that the signal-to-noise ratio is at most order one for the Wino mass
of a few hundred GeV.  We have also checked that the upward muon
events expected at the KM3NeT detector~\cite{848444}, assuming the
effective area of $1\,{\rm km^2}$ and taking account of the energy
loss of muons~\cite{hep-ph/0012350}, provide similar sensitivities to
the DeepCore.

The Wino DM annihilation may leave characteristic signatures on
astrophysical observations.  Gamma-ray observations by Fermi-LAT and
HESS severely restrict the DM annihilation cross section (see, e.g.,
Refs.~\cite{arXiv:1108.3546,arXiv:1110.6151} for recent works).  The
non-observations of DM-induced gamma-rays from dwarf galaxies excludes
the Wino mass below $\sim 400$\,GeV~\cite{arXiv:1108.3546}, although
there are astrophysical uncertainties.  On the other hand, the
cosmic-ray positron excess observed by PAMELA
satellite~\cite{arXiv:0810.4995} may be explained by the Wino DM
annihilation with mass of
200\,GeV~\cite{arXiv:0810.1892,arXiv:0812.4555} although it may
confront the constraints from gamma-rays and anti-protons.  The
observations of light element abundances also give stringent bound on
the DM annihilation cross section so as not to destroy light elements
during Big-Bang Nucleosynthesis (BBN).  It gives a lower bound on the
Wino mass as $m_{\tilde W} \gtrsim
200$\,GeV~\cite{astro-ph/0405583,arXiv:0810.1892}.  It may be
encouraging that the cosmic lithium problem may be solved for the Wino
mass of around this bound, which simultaneously may explain the PAMELA
anomaly.  DM annihilation also affects the recombination history of
the Universe, which results in the modification on the cosmic
microwave background (CMB)
anisotropy~\cite{arXiv:0905.0003,arXiv:0906.1197,arXiv:0907.3985}.
The constraint is comparable to that from BBN.  Taking these
constraints into account, we conservatively consider that the Wino must
be heavier than $\sim 200$\,GeV if it is the dominant component of DM.


Finally, we comment on a possibility of discovering a signal of AMSB
model at the LHC.  If we adopt the AMSB mass relation among gauginos,
gluino becomes relatively heavy.  Then, colored superparticles are
hardly produced at the LHC.  Thus, we focus on the detection of a Wino
signal.

If the neutral Wino $\tilde{W}^0$ is the LSP, we have a chance to
observe the track of charged Wino $\tilde{W}^\pm$ \cite{Feng:1999fu,
  hep-ph/0610277}.  This is because the mass difference between
charged and neutral Winos is so small ($\sim 160\ {\rm MeV}$) that the
decay length of $\tilde{W}^\pm$ becomes macroscopic
($c\tau_{\tilde{W}^\pm}\simeq 5\ {\rm cm}$).  Some of the produced
charged Winos may travel through several layers of inner trackers and
their track may be reconstructed.  In the ATLAS experiment, for
example, the charged Wino track can be reconstructed with almost 100
\% efficiency if $\tilde{W}^\pm$ hits the 3rd layer of the
semiconducter tracker (SCT) before it decays \cite{arXiv:0807.4987}.
Then, because of the smallness of $c\tau_{\tilde{W}^\pm}$ compared to
the detector size, $\tilde{W}^\pm$ decays before going through the
whole detector.  Such a charged Wino is identified as a high $p_T$
track which disappears in the middle of the detector.  Such a signal
does not exist in the SM, and hence is a smoking gun evidence of the
production of $\tilde{W}^\pm$.

The Wino pair can be produced by the Drell-Yan process at the LHC.
However, there is no high $p_T$ jet nor track in the final state in
such an event, and hence the event cannot be recorded.  In order to
trigger on the Wino production events, one can use the event with high
$p_T$ jet; such a jet can be from the initial state radiation.  Then,
at the parton level, the Wino production processes relevant for the
present study are the following:
\begin{eqnarray*}
  &&
  q\bar{q} \rightarrow \tilde{W}^+ \tilde{W}^- g,~~~
  gq \rightarrow \tilde{W}^+ \tilde{W}^- q,~~~
  g\bar{q} \rightarrow \tilde{W}^+ \tilde{W}^- \bar{q},
  \\ &&
  q\bar{q}' \rightarrow \tilde{W}^\pm \tilde{W}^0 g,~~~
  gq \rightarrow \tilde{W}^\pm \tilde{W}^0 q',~~~
  g\bar{q} \rightarrow \tilde{W}^\pm \tilde{W}^0 \bar{q}'.
\end{eqnarray*}
We calculate the cross section of the process
$pp\rightarrow\tilde{W}\tilde{W}j$; we perform the parton level
calculation, and we approximate the $p_T$ of jet by that of
final-state quark or gluon.  In the calculation of the cross section,
the helicity amplitude package HELAS \cite{Murayama:1992gi} and the
CT10 parton distribution functions \cite{Lai:2010vv} are used.  For
the phase space integration, we use the BASES package \cite{BASES}.
In the calculation, we require that the transverse momentum of the jet
be larger than 170, 270, and 370 GeV, and that at least one charged
Wino travels more than 44.3 cm which is the distance to the 3rd layer
of SCT from the beam pipe in the ATLAS detector \cite{Aad:2008zzm}.

\begin{figure}[tbp]
\begin{center}
\includegraphics[scale=1.7]{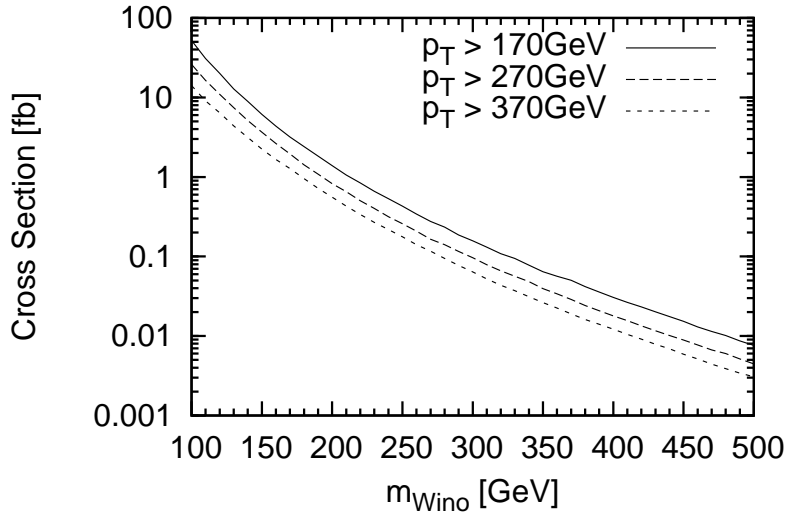}
\caption{
  Cross section for the process $pp\rightarrow\tilde{W}\tilde{W}j$ (with
  $j=q$ or $g$), for $\sqrt{s}=14\ {\rm TeV}$.  The transverse
  momentum of $j$ is required to be larger than 170, 270, and 370 GeV
  from above. }
\label{fig:cs}
\end{center}
\end{figure}

In Fig.\ \ref{fig:cs}, the cross section is plotted as a function of
the Wino mass.  In the high luminosity run with ${\cal L}=2\times
10^{33}\ {\rm cm^{-2}s^{-1}}$, for example, the so-called j370 trigger
is planned to be available, which requires a jet with $p_T>370\ {\rm
  GeV}$ \cite{Aad:2008zzm}.  Then, requiring 10 events with $p_T>370\
{\rm GeV}$ for the discovery, for example, Wino mass smaller than
$270\ {\rm GeV}$ ($330\ {\rm GeV}$) is covered by the LHC with the
luminosity of ${\cal L}=100\ {\rm fb}^{-1}$ ($300\ {\rm fb}^{-1}$),
where we have assumed that the background is negligible.  Thus, in
particular when $\mu$ is large, the LHC experiment still have a chance
to cover the parameter region which has not been excluded yet by the
current direct and indirect DM searches.  If the $p_T$ of the jet for
the trigger can be reduced, the LHC can cover the region with larger
Wino mass.

So far, we have assumed the pure AMSB relation among gaugino masses.
However, as we have mentioned, such a relation may be largely affected
by the Higgs and Higgsino loop diagrams.  With such an effect, the
gluino mass may become $\sim 1\ {\rm TeV}$ even when the Wino mass is
a few GeV.  In such a case, the conventional procedures of the SUSY
search using the missing energy distribution may work.

In summary, motivated by the recent report on the Higgs searches at
the LHC, which indicated excesses of Higgs-like events at around
$m_h\simeq 125\ {\rm GeV}$, we have investigated prospects for
confirmation of the AMSB scenario, particularly the detection of Wino
LSP.  We have considered the situation that the scalars except for
gauginos and Higgsinos are heavy enough so that they cannot be
produced at colliders.  Even in this unfortunate case, the Wino DM may
be detected through direct/indirect detection experiments.  Direct
detection efficiency crucially depends on the Higgsino mass, and if
the Wino and Higgsino masses happen to be close, future experiments
may find their signals.  The neutrino telescopes such as IceCube
DeepCore and KM3NeT also have a potential to discover the Wino LSP
through the observation muon and/or shower events induced by
high-energy neutrinos from DM annihilation at GC.

\section*{Acknowledgment}

We would like to thank M.~Ibe and T.~T.~Yanagida for useful
discussion.  This work is supported by Grant-in-Aid for Scientific
research from the Ministry of Education, Science, Sports, and Culture
(MEXT), Japan, No.\ 22244021 (T.M.), No.\ 22540263 (T.M.), No.\
23104001 (T.M.), No.\ 21111006 (K.N.), and No.\ 22244030 (K.N.).




\end{document}